\documentclass[%
 reprint,
 superscriptaddress,
 amsmath,amssymb,
 aps,
 prl,
]{revtex4-2}

\usepackage{graphicx}
\usepackage{dcolumn}
\usepackage{bm}
\usepackage{comment}
\usepackage{siunitx}

\newcommand*{\rom}[1]{\expandafter\@slowromancap\romannumeral #1@}
\def\BigRoman{\uppercase\expandafter{\romannumeral\number\count 255 }}
\def\Romannumeral{\afterassignment\BigRoman\count255=}

\begin{document}

\preprint{APS/123-QED}

\title{X-ray Microscopy Study of Freezing Sessile Droplets}

\author{Jae Kwan Im}
\affiliation{Department of Physics, Ulsan National Institute of Science and Technology, Ulsan, Republic of Korea}

\author{Hyeonjun An}
\affiliation{Department of Physics, Ulsan National Institute of Science and Technology, Ulsan, Republic of Korea}

\author{Seob-Gu Kim}
\affiliation{Pohang Accelerator Laboratory, Pohang University of Science and Technology, Pohang, Republic of Korea}

\author{Jae-Hong Lim}
\affiliation{Pohang Accelerator Laboratory, Pohang University of Science and Technology, Pohang, Republic of Korea}

\author{Joonwoo Jeong}
\email{jjeong@unist.ac.kr}
\affiliation{Department of Physics, Ulsan National Institute of Science and Technology, Ulsan, Republic of Korea}
\affiliation{UNIST Research Center for Soft and Living Matter, Ulsan National Institute of Science and Technology, Ulsan, Republic of Korea}

\date{\today}

\begin{abstract}
A sessile water droplet on a cold substrate freezes into a shape with a sharp apex because of water's expansion upon freezing, yielding a universal tip angle across various conditions. Using \textit{in situ} X-ray imaging, we report that this angle changes with substrate temperature, and the deviation originates from bubble formation during freezing. Three-dimensional tomography enables direct quantification of the effective ice-water density ratio, accounting for trapped bubbles. Incorporating this effective density ratio reconciles the temperature-dependent tip angles. We also confirm that a bubble-free frozen droplet in a vacuum chamber exhibits the universal tip angle. Furthermore, X-ray imaging allows us to measure the three-phase boundary angles \textit{in situ}, thereby validating the geometric theory behind tip formation. These findings advance our understanding of the freezing dynamics associated with multiphase systems and highlight the capabilities of high-resolution X-ray imaging in ice research.

\end{abstract}

\maketitle


The freezing of water is a ubiquitous phenomenon with significant implications across natural and engineered systems. 
Ice not only plays a central role in polar ecosystems, climate, and planetary processes~\cite{Snodgrass2017,Noble2020,Holland2022}, but also poses significant challenges to safety, reliability, and efficiency in technological contexts spanning from aviation and thermal management to food preservation and biomedical cryopreservation~\cite{Jaiswal2022,Rekuviene2024,Zhang2024,Mckenzie2024,Aarattuthodi2025,Tian2025}. 
The complex nature of freezing and melting stems from their multi-physics and multi-component characteristics. For example, coexisting, coupled gradients in density, temperature, and solute concentration can lead to fluid and material transport, as well as phase separation.
These effects manifest as gas inclusions~\cite{Carte1961,Madrazo2008,Hruba2018,Chu2019,Meijer2024,Shao2025}, brine channels, and residual liquid films~\cite{singha2018,Kamm2023,Chu2024,Kharal2024,Zhu2024,Gao2025} in natural ice having complicated internal structures.

The sharp tip formed by a frozen water droplet on a cold plate~\cite{Anderson1996,Ajaev2003,Snoeijer2012,Enriquez2012,Marin2014} is one of the most fascinating freezing phenomena that even pure water can exhibit, including ice spikes and snowflakes~\cite{Hill2004,Libbrecht2005}.
The surface-tension-defying formation of the tip singularity has been attributed to the expansion of liquid upon solidification~\cite{Kolasinski2007}. 
The geometry of the freezing front---a spherical-cap shape that ends perpendicularly to the solid-air interface---is responsible for the formation of the pointy tip and its universal angle regardless of the substrate temperature or the contact angle on a substrate~\cite{Marin2014}.

Here, we employ synchrotron X-ray imaging to resolve the fast dynamics of the freezing process and the 3D internal structure of the frozen droplets, enabled by X-ray tomography. 
By overcoming the limited spatial and temporal resolution in previous studies~\cite{Snels2023,Wu2025}, we quantitatively characterize the geometry of the freezing front and the trapped bubbles, whose sizes and spatial distributions depend on the freezing rate.
These observations reconcile the slight mismatch between experiments and theory reported in Mar\'{i}n et al.~\cite{Marin2014} and propose a revised model that accounts for trapped bubbles and the resulting substrate-temperature-dependent tip angles.

\begin{figure*}[t]
    \centering
    \includegraphics{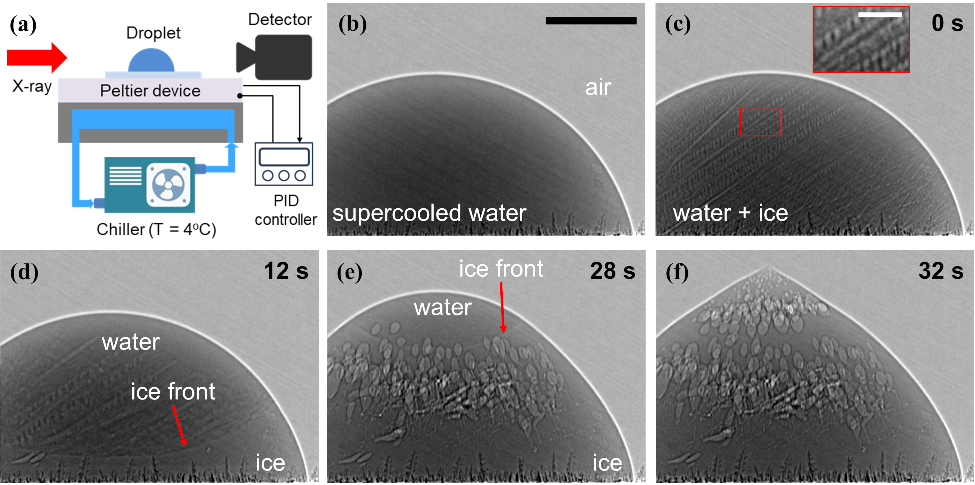}
    \caption{\label{fig1}
X-ray imaging of the freezing process of a sessile water droplet placed on the cold polycarbonate plate. After placing a droplet of \SI{2}{\micro\liter} in volume on the substrate at room temperature, we decrease the substrate temperature to \SI{-5}{\degreeCelsius}. (a) The sketch of the experimental setup. Four distinct stages of the droplet freezing are shown. (b) Liquid cooling. A droplet remains in the supercooled liquid phase despite sub-zero substrate temperature. The black scale bar is \SI{500}{\micro\meter}. (c) Recalescence. Dendritic ice nucleates; the inset with the white 100~\si{\micro\meter} scale bar displays its magnified view. (d) and (e) Freezing. Curved water(top)-ice(bottom) interface, water-ice-air boundary, and the trapped air bubbles are visible. (f) Solid cooling. The completely frozen droplet forms the pointy tip at the top. Elapsed time after the recalescence is shown in the top-right corner.}
\end{figure*}

\begin{figure*}[t]
    \centering
    \includegraphics{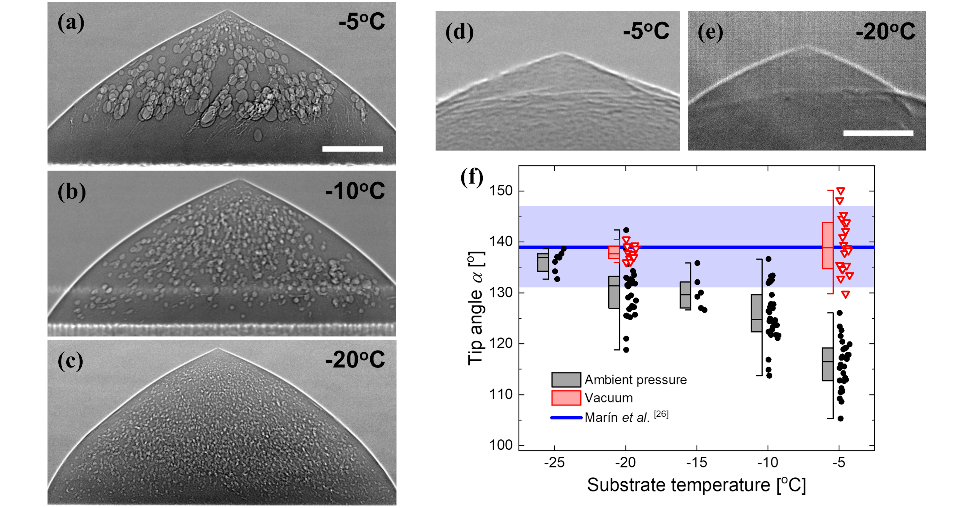}
    \caption{\label{fig2}
Effect of trapped bubbles on the tip angle according to the substrate temperature. 
The representative 2D X-ray images of the frozen water droplets of \SI{1}{\micro\liter} in volume on a cover glass at (a) \SI{-5}{\degreeCelsius}, (b) \SI{-10}{\degreeCelsius}, and (c) \SI{-20}{\degreeCelsius}, respectively, in the ambient condition. 
The representative images of the bubble-free frozen water droplet on a cover glass at (d) \SI{-5}{\degreeCelsius} and (e) \SI{-20}{\degreeCelsius}, respectively, in a vacuum chamber.
All scale bars are \SI{200}{\micro\meter}.  
(f) Measured tip angles depending on substrate temperature and presence of bubbles.
The black-filled circles and red-empty triangles represent the tip angles of individual frozen droplets with bubbles and without bubbles, respectively. 
The center line of the box plot corresponds to the median of each dataset, and the top and bottom lines are the 75th and 25th percentiles, respectively.
Blue solid line at 139 deg and the band around it indicate the average and standard deviation of the tip angles reported in Mar\'{i}n et al.~\cite{Marin2014}: $139\pm8$\si{\degree}.
    }
\end{figure*}

X-ray imaging visualizes the internal structure of a freezing water droplet with a pointy tip. 
The whole freezing process of a water droplet on a cold plate can be divided into four distinct stages based on its temperature transition characteristics: (1) liquid cooling (supercooling), (2) recalescence (nucleation), (3) freezing with an ice front advancing from the cold substrate, and (4) solid cooling \cite{Hindmarsh2003,Chaudhary2014,Wang2021}. 
In the liquid cooling stage shown in Fig.~\ref{fig1}(b), the water droplet remains in the supercooled liquid phase, and its temperature decreases below \SI{0}{\degreeCelsius}. 
Figure~\ref{fig1}(c) captures the moment of recalescence, which typically lasts tens of milliseconds~\cite{Chaudhary2014, Wang2021}. 
When a supercooled droplet forms ice nuclei and releases the latent heat, its temperature rises to \SI{0}{\degreeCelsius}.
Then, as shown in Figs.~\ref{fig1}(d) and~\ref{fig1}(e), the ice grows from the bottom.
In our X-ray observation, the three-phase boundary between ice, water, and air, as well as the formation of the pointy tip, is clearly visible. 
Notably, while the pointy tip is known to adopt a universal angle of $139\pm8$\si{\degree}~\cite{Marin2014}, regardless of the substrate's temperature and surface chemistry, we observe tip angles below this universal value at higher substrate temperatures. 
As plotted in Fig.~\ref{fig2}(f), the deviation from the universal value becomes more pronounced as the substrate temperature increases, with the tip angle reaching approximately \SI{120}{\degree} at \SI{-5}{\degreeCelsius}.

To reconcile this deviation in the tip angle, we focus on the temperature-dependent trapping of air bubbles during freezing.
Regardless of whether a prior degassing process has been applied, air bubbles are present in all the ice droplets we observed.
We presume that the small sessile droplets, with a volume of \SI{1}{\micro\liter} and a high surface-to-volume ratio, quickly dissolve the ambient gases during the freezing process after we place a droplet and cool the substrate. 
Figures~\ref{fig2}(a), (b), and (c) show that the spatial distribution and size of trapped bubbles strongly depend on the substrate temperature. 
See Supplemental Material for Supplemental Movies indicating that the temperature-dependent freezing rate, i.e., the advancing speed of the ice front, affects the bubble distribution and size.
In the case of a rapidly freezing droplet at \SI{-20}{\degreeCelsius}, the tiny bubbles of a relatively uniform size distribution, $17.1\pm4.8$~\si{\micro\meter}, are evenly distributed. 
In contrast, at \SI{-5}{\degreeCelsius}, resulting in slower ice-front advance, larger and polydisperse bubbles are concentrated near the top of the droplet. 
We propose that the fast-moving ice front immediately captures the gas bubbles extracted from the interface, whereas buoyant bubbles can outrun the slow-moving front and coalesce. Note that the ice front accelerates because the remaining liquid volume decreases as it advances in the sessile droplet geometry.

The tip angle of bubble-free ice matches the reported universal value, supporting our hypothesis that the large bubbles near the apex are responsible for the reduced tip angle. 
To observe the bubble-free ice, we repeat cycles of freezing and thawing in a vacuum chamber installed on X-ray imaging beamline. 
The resulting bubble-free ice, as shown in Figs.~\ref {fig2}(d) and~\ref {fig2}(e), recovers the universal tip angle, regardless of the substrate temperature.

\begin{figure*}[t]
    \centering
    \includegraphics{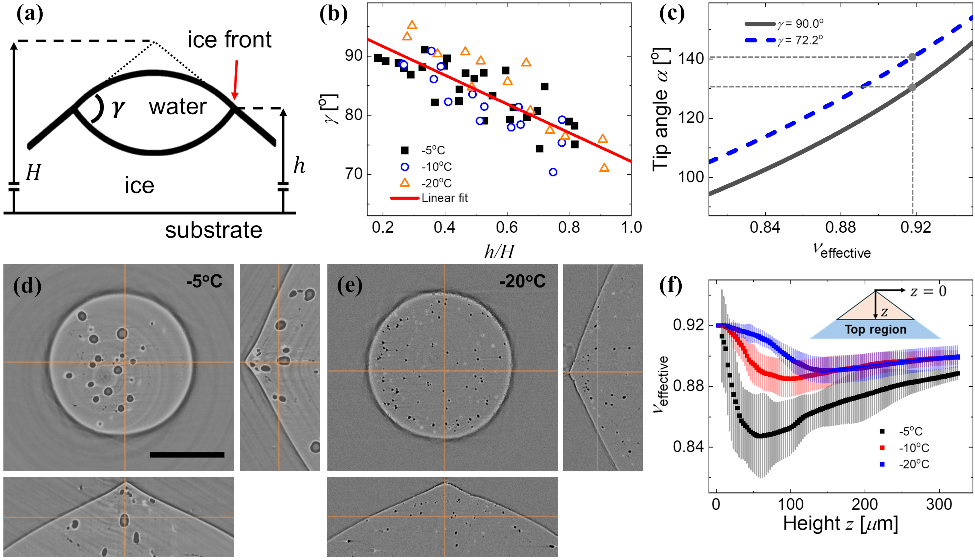}
    \caption{\label{fig3}
X-ray imaging-enabled characterization of droplet freezing.
(a) The schematic of the three-phase boundary. The advancing ice front has the height $h$ and the angle $\gamma$ to the air-water interface at the three-phase boundary. $H$ indicates the final height of the frozen droplet. 
(b) $\gamma$ as a function of the normalized height $h/H$ according to the substrate temperature. The data points are acquired from multiple droplets of the same volume in ambient conditions.
(c) Theoretical curves predicting the tip angle according to $\gamma$ and the effective ice-to-water density ratio $\nu_{\text{effective}}$. The black solid is for $\gamma$ = \SI{90}{\degree}, and the blue dashed line is for $\gamma$ = \SI{72.2}{\degree}, which is the estimated $\gamma$ at $h/H = 1$ by the linear fit in (b).
The representative X-ray tomographic cross-sections of the frozen water droplets of \SI{1}{\micro\liter} in volume on a cover glass at (d) \SI{-5}{\degreeCelsius} and (e) \SI{-20}{\degreeCelsius}, respectively, in the ambient condition. The scale bar is \SI{200}{\micro\meter}.
(f) Estimated $\nu_{\text{effective}}$ in the bubble-laden ice volume truncated at a distance $z$ from the end of the tip. The data points represent the average value of 20, 18, and 11 specimens at -5, -10, and \SI{-20}{\degreeCelsius}, respectively, and the standard deviations are shown as bands. 
    }
\end{figure*}

The theoretical model presented by Mar\'{i}n et al.~\cite{Marin2014} successfully explains the tip angles of bubble-laden ice by combining our X-ray imaging measurements of the three-phase boundary with the \textit{effective} ice-to-water density ratio.
First, we directly measure $\gamma$, the angle between air-water and water-ice interfaces, and find that $\gamma$ decreases from \SI{90}{\degree} as the ice front advances, down to $72.2\pm 2.4$\si{\degree} in the vicinity of the tip. 
Fig.~\ref{fig3}(b) shows that this decrease occurs regardless of the substrate temperatures. 
Then, this newly measured $\gamma$ updates the model prediction, $f(\gamma-\theta)+f(\theta)=\nu[f(\gamma-\theta)+\frac{\pi}{3}\tan\theta]$ with $f(\theta)=\frac{\pi}{3} (\frac{2-3\cos(\theta)+\cos^3(\theta)}{\sin^3(\theta)})$ and $\rho_{\mathrm{ice}}/\rho_{\mathrm{water}} \equiv \nu = 0.92$, to give the tip angle $\alpha = \SI{141}{\degree}$. 
This $\alpha$ provides a better fit for both their experimental observation, $\alpha=139\pm8 \si{\degree}$, and our measured value $\alpha = 136\pm2 \si{\degree}$ at \SI{-25}{\degreeCelsius} compared to the previous calculated value of \SI{131}{\degree} assuming $\gamma \approx \SI{90}{\degree}$~\cite{Marin2014}.
Note that experimental and theoretical studies have suggested that $\gamma$ deviates from the ideal conduction-only \SI{90}{\degree} in consideration of ice as a thermal resistor between water and cold substrate and convection in the surrounding air \cite{Tembely2019, Zhang2019, Zhang2020, Miao2024}. Our measurement provides direct, measurement-based evidence that $\gamma$ decreases as ice thickens.

In addition to $\gamma<\SI{90}{\degree}$, the effective ice-to-water density ratio $\nu_{\mathrm{effective}}$, smaller than 0.92 because of the trapped bubbles, justifies the substrate temperature-dependent deviation of $\alpha$ from the universal value.
The trapped bubbles can reduce the effective density of the ice, thus influencing the tip angle as described by the theoretical model~\cite{Li2022}.  
Figure~\ref{fig3}(c) shows the model prediction of tip angles as a function of $\nu_{\mathrm{effective}}$, which is smaller than the universal value based on $\nu=0.92$ and $\gamma=\SI{90}{\degree}$.
From the 3D tomography of the apex region of ice, as shown in Figs.~\ref{fig3}(d) and~\ref{fig3}(e), we segment and measure the volumes of the ice and bubble regions, respectively.
Under the mass conservation assumption, the effective density ratio $\nu_{\mathrm{effective}}$ corresponds to $\rho_{\mathrm{ice+bubble}} / \rho_{\mathrm{water}}=V_{\mathrm{water}} / V_{\mathrm{ice+bubble}}$, where $V_{\mathrm{water}}$ is $\nu=0.92$ times the measured $V_{\mathrm{ice}}$, and $V_{\mathrm{ice+bubble}}$ is also measured from the tomography.
Figure~\ref{fig3}(f) plots $\nu_{\mathrm{effective}}$ in the apex region above $z$. 
The warmer substrate results in a greater reduction in $\nu_{\mathrm{effective}}$ in the apex region, and consequently smaller $\alpha$. 
The higher the temperature of the substrate, the greater the change in the value of $\nu_{\mathrm{effective}}$.
This results from larger, polydisperse bubbles concentrated in the top region, which outrun the slow ice front and coalesce during floatation.
For instance, the top region of the pointy ice frozen on a substrate at \SI{-5}{\degreeCelsius} exhibits $\nu_{\mathrm{effective}}$ as low as 0.85. 
In the model, this small $\nu$ leads to $\alpha \approx \SI{117}{\degree}$ with $\gamma = 72.2\si{\degree}$, which is similar to our experimentally measured $\alpha = 116\pm5\si{\degree}$. However, the narrow volume of the sharp tip region cannot accommodate large bubbles, causing $\nu$ to vary in the top region. Therefore, how $\nu_{\mathrm{effective}}(z)$ determines $\alpha$ warrants further investigation.

In conclusion, our high-resolution \textit{in situ} X-ray imaging elucidates the internal structure and freezing dynamics of sessile water droplets on cold substrates. 
Our results reveal that the tip angle of a frozen droplet is significantly modulated by the presence of air bubbles trapped within the ice. 
We demonstrate that slower freezing rates, associated with higher substrate temperatures, lead to the formation of larger trapped bubbles that effectively reduce the ice-water density ratio, thereby decreasing the tip angle from its theoretical universal value.
This hypothesis is corroborated by 3D tomographic quantification of bubble volume and experiments with bubble-free droplets, which restore the universal tip angle. 
Moreover, detailed measurements of the freezing front geometry refine theoretical predictions, showing that the angle at the ice-air interface deviates from \SI{90}{\degree}. 
Together, these findings extend our understanding of multiphase freezing phenomena by highlighting the critical role of trapped gas inclusions in determining the morphological evolution of freezing droplets. 
They also showcase the capability of high-resolution X-ray imaging techniques to probe complex internal structures in dynamic phase transitions.

\textit{X-ray imaging and analysis:}~X-ray imaging experiments were performed at the beamline 6C Bio Medical Imaging of the Pohang Light Source-\Romannumeral 2~(PLS-\Romannumeral 2). 
The X-ray energy was 12-14 keV, and the imaging pixel resolution was \SI{0.65}{\micro\meter}.
The time interval between sample placement and the first image acquisition was approximately 30 seconds due to a safety procedure involving hutch evacuation and shutter opening. 
The X-ray exposure time for a typical snapshot was 100 ms, and the freezing process was recorded at 1 frame per second. 
Flat-field correction was applied using the background and dark images~\cite{Seibert1998}. 
The background image was taken without a sample under identical conditions, and the dark image recorded the detector noise level with no X-ray illumination. 
To calculate the X-ray transmittance image, the dark image was subtracted from both the sample and background images, and the corrected sample image was then divided by the corrected background image. 
The contrast of the images shown in the figures is adjusted to improve interfacial visibility. 
For tomography of a frozen droplet, the sample stage was rotated by \SI{180}{\degree}, and 360 projection images were collected at \SI{0.5}{\degree} intervals over 8 seconds.
Three-dimensional reconstruction was performed using Octopus 8.7 software (XRE, Gent, Belgium).
From each 3D tomogram, ice and bubble regions were segmented using the Statistical Region Merging method in ImageJ~\cite{Nock2004}, and voxel counts were used to determine their volumes.

\textit{Sample preparation:}~For all experiments, we used deionized water with a resistivity of \SI{18.2}{\mega \ohm \cm}. 
Two different substrates were used as received: a cover glass (Duran, DWK Life Sciences, Germany) and a polycarbonate plate (ArtRyx, South Korea).
The substrate was placed on a Peltier device connected to a PID controller and a chiller. Its temperature was controlled from room temperature to \SI{-25}{\degreeCelsius} under ambient conditions: $24\pm0.4~\si{\degreeCelsius}$ and relative humidity of $30\pm2\%$.
1 or 2~\si{\micro\liter} of deionized water was manually deposited on the substrate at room temperature. 
Then, we started the X-ray imaging and cooled the Peltier device to a target temperature at a rate of \SI{-2}{\degreeCelsius / \second}.
Note that after the substrate reached the target temperature, supercooled water droplets typically exhibited recalescence within 30 seconds.
To produce bubble-free ice, we employed a freeze-thaw method~\cite{Ninham2020} in a custom-made vacuum chamber at approximately 10 kPa. 
We typically applied four freeze-thaw cycles before creating the final bubble-free ice.
In the special case of \SI{-5}{\degreeCelsius} in the vacuum chamber, we added 4 ppm of ice-nucleation protein (Snomax, U.S.A.) to water to facilitate freezing from the supercooled state; without ice-nucleation protein, we were unable to freeze the supercooled droplet before complete evaporation.

\begin{acknowledgments}
This work was supported by Korea Polar Research Institute (KOPRI) grant funded by the Ministry of Oceans and Fisheries (KOPRI PE25900). The authors also acknowledge partial financial support from the Korean National Research Foundation through Grant No. RS-2024-00437443. Experiments using PLS-II were supported in part by the Ministry of Science and ICT (MSIT, RS-2022-00164805, Accelerator Application Support Project) and POSTECH.
\end{acknowledgments}

%

\end{document}